\documentclass{elsart}

\usepackage{natbib}

\usepackage{graphicx}

\usepackage{amssymb}

\begin{document}

\begin{frontmatter}



\title{Transient conditions for biogenesis on low-mass exoplanets with escaping hydrogen atmospheres}


\author[LMD]{R. Wordsworth\thanksref{CHI}}

\address[LMD]{Laboratoire de M\'et\'erologie Dynamique, Institut Pierre Simon Laplace, Paris, France.}
\thanks[CHI]{Current address: Department of the Geological Sciences, University of Chicago, 60637 IL, USA}



\begin{abstract}
Exoplanets with lower equilibrium temperatures than Earth and primordial hydrogen atmospheres that evaporate after formation should pass through transient periods where oceans can form on their surfaces, as liquid water can form below a few thousand bar pressure and H$_2$-H$_2$ collision-induced absorption provides significant greenhouse warming. The duration of the transient period depends on the planet size, starting H$_2$ inventory and star type, with the longest periods typically occurring for planets around M-class stars.  As pre-biotic compounds readily form in the reducing chemistry of hydrogen-rich atmospheres, conditions on these planets could be favourable to the emergence of life. The ultimate fate of any emergent organisms under such conditions would depend on their ability to adapt to (or modify) their gradually cooling environment. 
\end{abstract}

\begin{keyword}
Prebiotic environments; Atmospheres, evolution; Terrestrial planets; Extrasolar planets; Solar radiation
\end{keyword}

\end{frontmatter}

\section{Introduction}

Recent developments in exoplanet characterization \cite[e.g., ][]{Tinetti2007,Knutson2007,Bean2010} have raised the possibility that in the relatively near future, we may be able to search for biosignatures in the atmospheres of terrestrial-mass planets. However, as Earth is the only current example of a life-supporting world that we know of, there is still no consensus as to the form such biosignatures will take. More fundamentally, there are still major unknowns as to the prevalence of conditions necessary to create and sustain life on other planets in general.

To date, much of the discussion has centered on the concept of the habitability zone - the orbital range around a star in which temperatures are in the right range for a planet to have surface liquid water. The standard definition of this region \citep{Kasting1993} is based on the assumption that a habitable planet will have a climate dominated by the greenhouse warming of CO$_2$ and H$_2$O, with temperature regulation provided by the carbonate-silicate cycle via plate tectonics. Alternative hypotheses exist: for example, \cite{Stevenson1999} proposed that if they retain primordial H$_2$-He envelopes, planetary embryos ejected from their systems during planet formation could be kept warm enough for surface liquid water even as they drift through interstellar space. Following this, \cite{Pierrehumbert2011} noted that orbitally bound exoplanets far from their host stars might also be habitable to life if they form with relatively thin hydrogen envelopes and manage to preserve them against escape. Both these studies suggest that a definition of the habitable zone based on H$_2$O-CO$_2$ warming only may be too narrow. However, the percentage of planets left with exactly the right atmospheric pressures to allow habitable temperatures after the early stages of atmospheric erosion is likely to be small, and the observation of rocky planets that are free-floating or in distant orbits around other stars is technically challenging.

Here the impact of hydrogen greenhouse warming is considered for exoplanets that are closer-in and hence may lose their hydrogen atmospheres entirely. Major uncertainties remain in planetary formation theory, but it is generally understood that planets of between 1 and 20 Earth masses can form with primordial hydrogen envelopes either due to capture of the gas from the solar nebula \citep{Rafikov2006} or outgassing during accretion \citep{ElkinsTanton2008}. In the case of Earth and Venus, these envelopes may have been lost to space extremely rapidly, while as noted by \cite{Stevenson1999} and \cite{Pierrehumbert2011}, planets that are in distant orbits or are ejected to interstellar space during formation can retain them permanently. In the intermediate situations where hydrogen atmospheres persist some time after formation but are eventually removed, the planet's climate will clearly evolve over time. Here I argue that during this evolution, transient periods will occur where water exists as a liquid on the planet's surface, and organic molecules may form in the atmosphere via photochemistry.

In the following section, the greenhouse effect of hydrogen atmospheres is discussed and an empirical expression for the pressure necessary to maintain surface temperatures in the correct range for liquid water is derived. Next, a simple assessment of atmospheric loss due to thermal escape is presented, and transient habitable times as a function of planet mass, orbital distance and starting hydrogen inventory are estimated. Finally, the effects of processes not included in the model are considered, along with the potential implications of the results.

\section{Analysis}
\subsection{Climate}

In atmospheres dominated by hydrogen, the main opacity source in the infrared is H$_2$-H$_2$ collision-induced absorption (CIA). Other gases can also contribute to infrared opacity and hence greenhouse warming, but only if they are present in sufficient amounts in the mid- to high atmosphere to absorb in the spectral regions where hydrogen becomes transparent. Helium can warm through He-H$_2$ or He-He CIA, but its effects should be secondary to those of hydrogen. In addition, it will not be present in large quantities if the atmosphere is created by accretion outgassing \citep{ElkinsTanton2008}. For temperatures of a few hundred Kelvin or less, the dominant equilibrium O-C-N species will be H$_2$O, CH$_4$ and NH$_3$ if hydrogen is the majority gas, as in the atmospheres of the Solar System gas giants \citep[e.g.][]{Moses2011}. The primary effect of these gases should be to increase greenhouse warming, although they can also influence the climate via cloud and haze formation (see Discussion).  For simplicity, the climatic effect of gases other than H$_2$ is neglected in the main part of this analysis, although some calculations involving H$_2$-H$_2$O are also discussed.

To allow rapid assessment of climate for a range of cases, one-dimensional radiative-convective calculations were used to produce an analytical expression for the globally averaged outgoing longwave radiation (OLR) $F_{out}$ and shortwave albedo $A$ assuming pure H$_2$ atmospheres. The model used was similar to that described in \cite{Wordsworth2010,Wordsworth2010b}:
there were 32 spectral bands in the longwave and 36 in the shortwave, and AD Leo and solar data \citep{Segura2005} were used for M-class and G-class stellar spectra, respectively. A two-stream scheme \citep{Toon1989} was used to compute fluxes, and the radiative effects of Rayleigh scattering in the visible were taken into account using the H$_2$ refractive index from \cite{Hecht2001}.

\begin{figure}[h]
	\begin{center}
		{\includegraphics[width=3.5in]{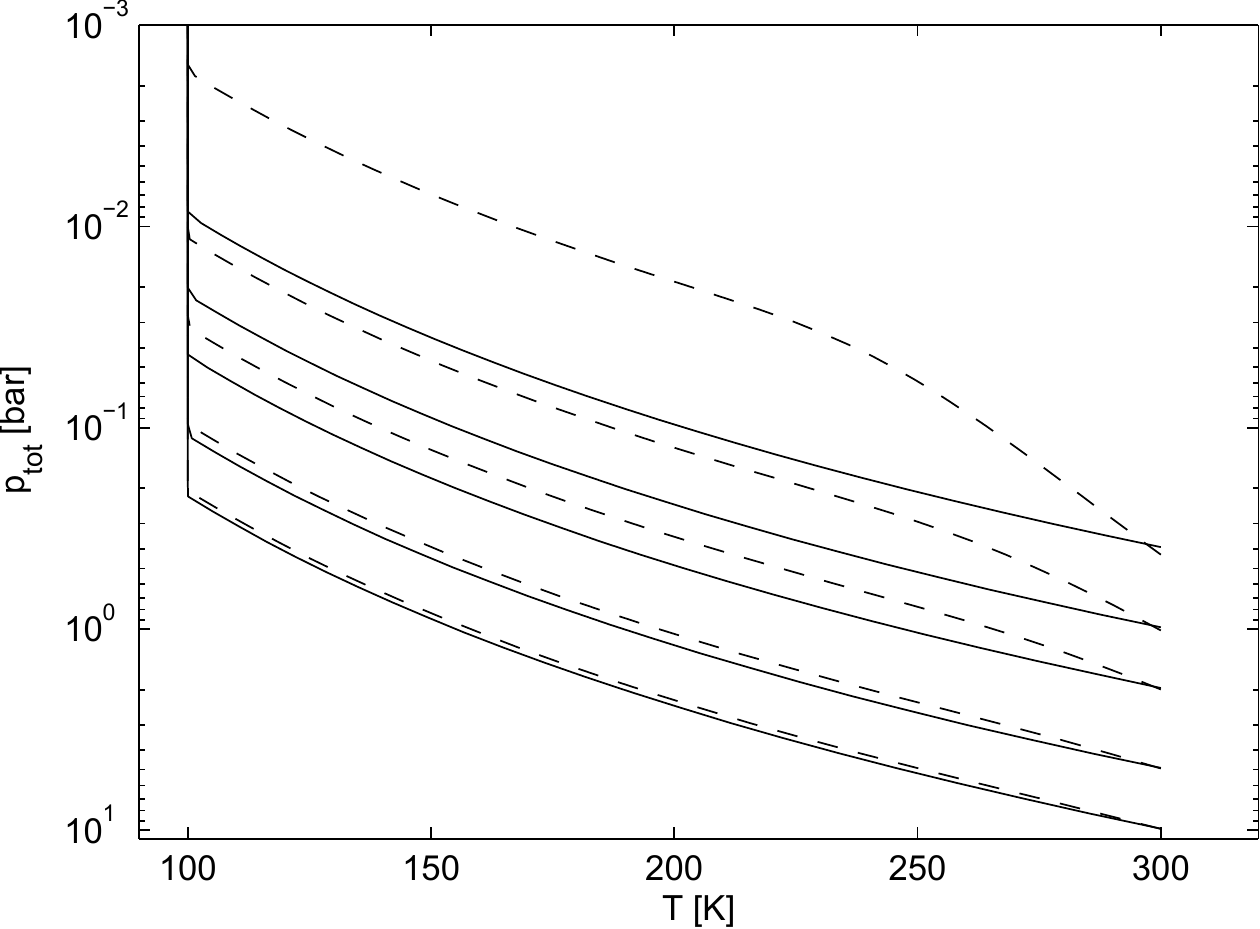}}
	\end{center}
	\caption{Example temperature profiles used in the climate calculations for pure H$_2$ atmospheres (solid lines) compared with those for mixed H$_2$-H$_2$O atmospheres (dashed lines).}\label{fig:dry_vs_wet_profs}
\end{figure}

Simulations were performed with specific heat capacity $c_p$ = 14.31 kJ kg$^{-1}$ K$^{-1}$ \citep[value at 300 K]{CRC2000} and molar mass $\mu_{H_2}= 2.02$ g to calculate $F_{out}$ as a function of surface pressure $p_s$ and gravity $g$. Hydrogen CIA opacities were calculated from tabulated spectral absorption data \citep{Borysow2002}. The vertical temperature profile was assumed to follow the dry ideal adiabat $T(z)=T_s - (g\slash c_p)z$, with the transition to an isothermal region (`stratosphere') at low pressures assumed to occur at 100 K (Figure \ref{fig:dry_vs_wet_profs}). Comparison with a time-stepping version of the model and simulations where the isothermal transition was neglected entirely showed that the stratospheric temperature did not have a significant influence on the value of $F_{out}$, in agreement with previous studies \citep{Pierrehumbert2011}.

\begin{figure}[h]
	\begin{center}
		{\includegraphics[width=4.5in]{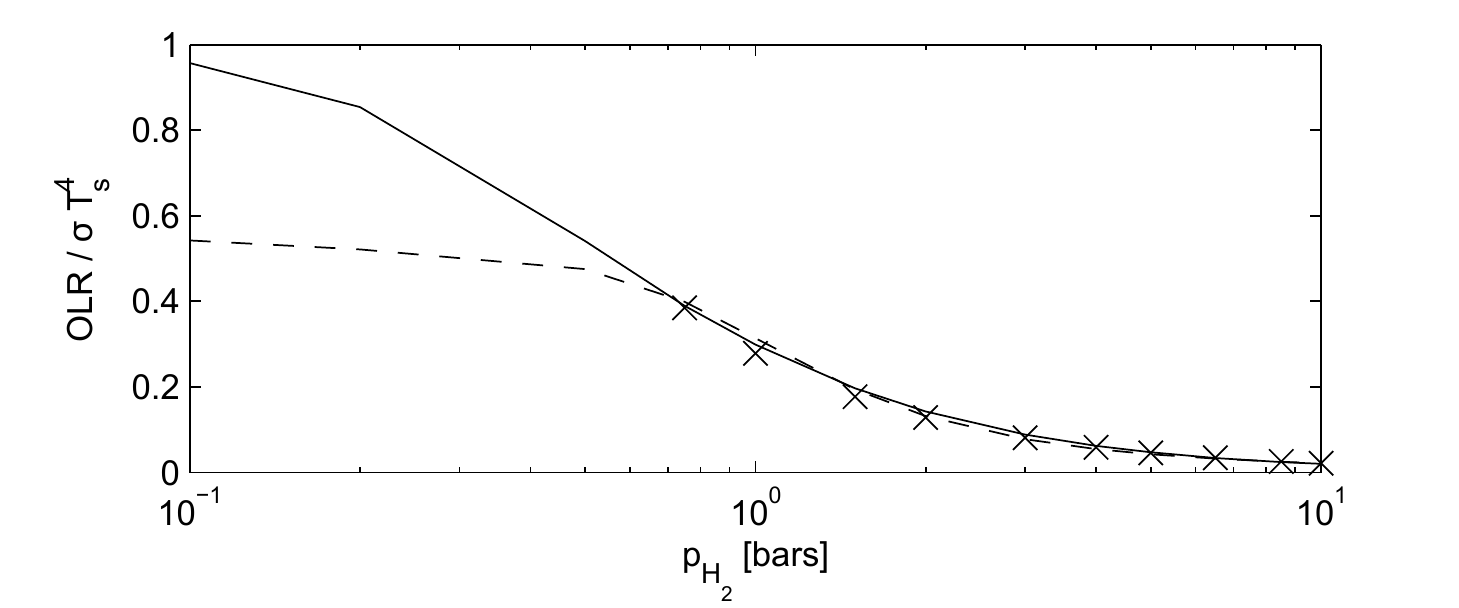}}
	\end{center}
	\caption{Calculated effective OLR ($F_{out} \slash \sigma T_s^4$) as a function of total H$_2$ pressure for pure H$_2$ (solid line) and mixed
H$_2$-H$_2$O (dashed line) atmospheres. As can be seen, significant differences between the two cases only occur below around $p_{H_2} = 0.5$ bar. The empirical fit to the pure H$_2$ curve given by (\ref{eq:OLRparam}) is shown as a series of crosses between 0.5 and 10 bar.}\label{fig:dry_vs_wet}
\end{figure}

We are most interested in greenhouse warming for surface temperatures in the range $T_{min}<T_s<T_{max}$, where $\{T_{min},T_{max}\} = \{ 273, 333\}$ K are conservative temperature limits for habitability to Earth-like microbial life. The variation of the effective OLR $F_{out} \slash \sigma T_s^4$ with temperature was found to be slow, so calculations were performed at $T_s = 300$ K only and the temperature dependence of the effective OLR was neglected. Given this simplification, the expression
\begin{equation}\label{eq:OLRparam}
F_{out} \slash \sigma T_s^4=a p_s^{-b} g_*^{b\slash 2}
\end{equation}
was calculated from the numerical results  within the limits $0.5 < p_s < 10$ bar and $10 < g < 30$ ms$^{-2}$ using a logarithmic least squares fit (see Figure \ref{fig:dry_vs_wet}). Here $p_s$ is in bars,  $a=0.279$, $b=1.123$, $g_* = g\slash g_{ref}$ and $g_{ref}=20.0$ ms$^{-2}$.

To determine the potential error in the analysis due to the assumption of pure H$_2$ atmospheres, a few additional calculations were performed including the presence of water vapour. In this case, saturation of H$_2$O vapour in the atmosphere was assumed, and the lapse rate was calculated with H$_2$O assumed to be an ideal gas. The effect of H$_2$O on the radiative transfer was included in a similar way to in previous studies \citep[e.g.][]{Wordsworth2010b}. High resolution spectra were computed from the HITRAN database \citep{Rothman2009} on a 6 $\times$ 8 $\times$ 8 temperature, pressure and H$_2$O volume mixing ratio grid with values $T = \{100, 150,\ldots, 350 \}$ K, $p  = \{ 10^{-2}, 10^{-1}, \ldots, 10^5 \} $ mbar and $q_{H_2O}=\{10^{-7}, 10^{-6}, \ldots, 10^{0} \}$, respectively, before conversion to a correlated-$k$ format for use in the radiative-convective model.

Figure \ref{fig:dry_vs_wet_profs} shows the temperature profiles vs. pressure for pure H$_2$ and \mbox{H$_2$-H$_2$O} atmospheres with constant stratospheric temperatures of 100 K. Figure \ref{fig:dry_vs_wet} shows the resulting effective OLR calculated using the radiative-convective model for a planet with $g=g_{ref}$. As can be seen, water vapour changes the lapse rate in the low atmosphere, primarily via latent heat release, but its effects decrease rapidly as the total pressure increases beyond a few fractions of a bar (the vapour pressure of H$_2$O is $\sim 0.035$ bar at 300~K). These lapse rate changes were more than compensated for climatically by increased infrared absorption - as shown by Figure  \ref{fig:dry_vs_wet}, water vapour generally decreased the effective OLR and hence increased the greenhouse effect. Nonetheless, its effects were not significant for $p_H{_2}$ values above about 0.4 bar. Hence equation (\ref{eq:OLRparam}) was judged an acceptable approximation for use in the following analysis.

In the visible, the radiative transfer in pure H$_2$ atmospheres is dominated by a combination of CIA and Rayleigh scattering. For surface pressures greater than around 20 bar, the fraction of starlight reaching the surface is small, and calculation of the surface temperature becomes difficult without knowledge of the planet's internal heat flux. In this analysis, all climate calculations were restricted to atmospheric pressures of 10 bar or less, for which it was sufficient to determine the planet's bond albedo in the visible. A surface albedo of 0.2 (typical for a rocky planet) was assumed. Albedos $A\simeq 0.25 + 0.15 $log$_{10}(p_s[\mbox{bar}])$ and $\simeq 0.15$ were found for G-class and M-class stars, respectively. The latter values were lower and approximately constant because the effects of Rayleigh scattering depend strongly on wavelength, and M-class stars have redder spectra. The variation of albedo with surface gravity was found to be small (under 0.025 at $T_s = 300$ K) and hence ignored. The effect of water vapour on albedo due to absorption was slightly greater (maximum decrease of $\sim 0.1$), but given the fact that it tended to increase warming and the inherent uncertainties due to haze and cloud cover, it was also neglected for simplicity. Finally, the reduction in visible flux early in the lifetime of G-class stars \cite[]{Gough1981} was taken into account, with the G-star stellar luminosity taken to be $L=0.7 L_\odot$. In the M-class case, the value $L=0.01205 L_\odot$ (representative of the red dwarf GJ581; \cite{vonBraun2011}) was used.

Planets with slowly escaping atmospheres will be in thermal equilibrium, $F_{out}=F_{in}$, if the loss timescale $t_{loss}$ is much longer than the radiative adjustment timescale $t_{rad}=c_p p_s T_s \slash g \sigma T_e^4$ \citep{GoodyYung1989}\footnote{Here we use $T_s$ rather than the smaller $T_e$ as an upper limit estimate of the mean atmospheric temperature.}, where $\sigma$ is the Stefan-Boltzmann constant, $g$ is gravity and $T_e$ is the characteristic emission temperature. Hydrogen has a mean optical depth of 1 in the infrared at around 0.2 bar pressure, so for e.g., a 20 bar atmosphere with surface temperature 300~K, the emission temperature should be $T_e = T_s (p_e \slash p_s)^{R \slash (c_p \mu_{H_2})} \sim 80$~K, where $R$ is the ideal gas constant. This yields $t_{rad}= 12,000$ years for an Earth-gravity planet. The presence of a primordial steam atmosphere could complicate this picture \cite[e.g. ][]{Valencia2008}, but only for planets that lose their entire hydrogen inventory in the first few million years after formation. In the following analysis this possibility is neglected and $F_{out}=F_{in}$ is assumed.

Writing $F_{in} = \frac 14 (1-A)(L\slash{4\pi d^2})$, where $L$ is the stellar luminosity and $d$ the orbital distance, and making the standard 1D modelling assumption of efficient horizontal heat transport by the atmosphere, we find
\begin{eqnarray}\label{eq:pressures}
p_s &=& \left( \frac{G}{\nu^2 g_{ref}} \right)^{1\slash 2} M^{1\slash 2-\beta} \left(\frac{ a\sigma T_s^4 }{(1-A)L\slash 16\pi d^2 } \right)^{1\slash b},\\
 &\propto&  M^{0.23} T_s^{3.56} d^{1.78}(1-A)^{-0.89}\nonumber
\end{eqnarray}
where the mass-radius relation $r = (\alpha r_\oplus \slash M_\oplus^\beta) M^\beta =\nu M ^\beta$  from \cite{Sotin2007} is used to write $g$ in terms of the planetary mass $M$. For simplicity the rocky planet values $\alpha=1.0$ and $\beta=0.274$ are assumed throughout the calculation. The approximately constant albedo in the M-star case allows direct calculation of $p_s[M,d,T_s]$, while in the G-star case, iteration is necessary for a self-consistent calculation. For an example 5$M_{\oplus}$ rocky planet at 2.5 AU around a Sun-like star with $L=0.7L_\odot$, iteration of (\ref{eq:pressures}) predicts an albedo of 0.35 and 4.3 bars of H$_2$ necessary to maintain a surface temperature of 300 K. The saturation temperature of H$_2$ at this pressure is less than 25 K \citep{CRC2000}, so atmospheric condensation does not occur at any altitude. As \cite{Stevenson1999} and \cite{Pierrehumbert2011} have already noted, hydrogen is an extremely potent, effectively incondensible greenhouse gas. Its main disadvantage in the context of achieving habitable temperatures is of course that usually, either far too much or too little of it is present.

\subsection{Atmospheric escape}

Unless hydrogen is continuously supplied via vulcanism or other surface processes at the same rate at which it escapes, habitable conditions due to H$_2$-H$_2$ warming will clearly be transient on exoplanets with escaping atmospheres. The habitable duration will depend primarily on the host star's extreme ultraviolet (XUV) emission and the planet's size and orbital distance. Erosion due to the stellar wind and meteorite impacts can also play a role in hydrogen loss from rocky planets, but thermal / hydrodynamic escape processes should dominate in the early phase of a system's evolution when stellar XUV levels are high \citep{Tian2005}.

For an atmosphere undergoing thermal escape, the flux of hydrogen to space depends primarily on the thermal escape parameter $\lambda=r \slash H_{exo}$, where $H_{exo}=k_B T_{exo} \slash m_H g$ is the scale height, $k_B$ is the Boltzmann constant, $T_{exo}$ is the temperature just below the exobase (exospheric temperature) and $m_H$ is the atomic mass of hydrogen\footnote{In the main part of this analysis efficient H$_2$ dissociation is assumed, so that the dominant escaping species is atomic hydrogen.}. The exospheric temperature is the most important unknown in $\lambda$. It may depend on factors such as the XUV heating, coupling with the lower atmosphere via thermal conduction and wave propagation, and cooling to space due to radiatively active molecules / ions such as H$_3^+$ \citep[e.g., ][]{Koskinen2007}. Here $T_{exo}$ is calculated using the approach of \cite{Lammer2003} as
\begin{equation}\label{eq:Texo}
T_{exo}^s=C F_{XUV} \slash g  + T_{min}^s,
\end{equation}
where $F_{XUV}$ is the stellar XUV flux, $T_{min}$ the temperature at the base of the thermosphere, $s=0.73$ and $C=1.67 \times 10^6$ K$^s$ W$^{-1}$ m$^3$ s$^{-2}$. In \cite{Lammer2003}, it was assumed following \cite{Bauer1971} that $T_{exo}$ is primarily determined by a balance between XUV energy input and energy loss to the lower atmosphere due to thermal conduction, with the latter  equal to $K(T_{exo})=K_0T_{exo}^s$ ergs K$^{-1}$ cm$^{-1}$ s$^{-1}$, where $K_0 = 16.4$ for atomic hydrogen. This approach neglects the possible effects of radiative cooling and internal heat sources, which we discuss later.

The temperature $T_{min}$ is included to account for conductive coupling between the thermosphere and the lower atmosphere. For all calculations assuming XUV-dominated escape, we set this parameter to the representative cold-trap temperature 100 K. However, the exact value of $T_{min}$ had a negligible influence on the results, as by the time the exospheric temperature drops to a few hundred Kelvin or lower, thermal escape rates are extremely small for planets of mass greater than $\sim 1~m_E$.

The stellar XUV flux depends primarily on the star type and age. Although XUV emissions may vary widely in a given stellar population (particularly for M-stars) the XUV flux tends to saturate at a value \mbox{$F_{XUV}\sim 10^{-3.1} L\slash{4\pi d^2}$} early in a star's lifetime, for a time $t_{XUV}$ that ranges from around 100~million years (G-stars) to 1~billion years (M-stars) \citep{Selsis2007}. After this it decays relatively rapidly according to a power law \citep{Ribas2005}. In most cases, the majority of atmospheric erosion therefore occurs early in a planet's lifetime. When other heating sources are negligible, an upper limit on atmospheric escape
\begin{eqnarray}\label{eq:elim}
\phi &=& \frac{\epsilon F_{XUV}}{4 r} \\\nonumber
     &=& \frac{\epsilon F_{XUV}}{4 \nu M^\beta}  \qquad \mbox{[Pa s$^{-1}$]}
\end{eqnarray}
where $\epsilon$ is an efficiency factor, can be derived. Equation (\ref{eq:elim}) assumes that the averaged XUV energy input to the atmosphere $F_{XUV} \slash 4$ ($\times$ the factor $\epsilon$) is used to overcome the gravitational potential energy of exospheric hydrogen atoms. For convenience when coupling with the climate equations, the hydrogen flux is converted to a loss rate of surface pressure per unit time. This makes use of the fact that surface pressure is equal to $m_{H_2} g$ times the area number density of H$_2$ molecules, and the potential energy of one hydrogen atom is $GMm_H\slash r$ (with $m_{H_2} \approx 2 m_H$).

This simple energy balance approach gives a reasonable estimate of escape rates for values of $\lambda$ equal to or less than around 2.8, for which the escape can be regarded as hydrodynamic \citep{Volkov2011}. For extremely small values of $\lambda$, the outer layers of the atmosphere extend to several planetary radii and the value of $\phi$ may be greater than that given in (\ref{eq:elim}). Hydrodynamic escape in such atmospheres is a formidable problem involving three-dimensional effects and coupled physics/chemistry, which we do not attempt to tackle here. As will be seen, the frequency of transient habitable conditions is most likely limited by the total amount of hydrogen that can be removed from the atmosphere, so the use of (\ref{eq:elim}) allows for a conservative estimate.

For $\lambda > 2.8$, the flux of hydrogen to space was estimated from the Jeans formula
\begin{equation}\label{eq:jeans}
\phi =  v_p n_{exo} f(\lambda) m_H g, \qquad \mbox{[Pa s$^{-1}$]}
\end{equation}
where $ v_p = (2k_B T_{exo} \slash m_H)^{1\slash 2}$ is the most probable particle speed, $n_{exo}=1\slash (\sigma_H H_{exo})$ is the particle number density at the exobase, \mbox{$\sigma_H=\pi (3.1 \times 10^{-12})^2$~m$^{2}$} is the hydrogen atom collision cross-section\footnote{Value at 1000 K, taken from \cite{Pierrehumbert2011BOOK}.} and $f(\lambda)=(1+\lambda)e^{-\lambda} \slash 2 \sqrt{\pi}$ is the fraction of upward-moving particles with speed greater than the escape velocity.

\begin{figure}[h]
	\begin{center}
		{\includegraphics[width=4in]{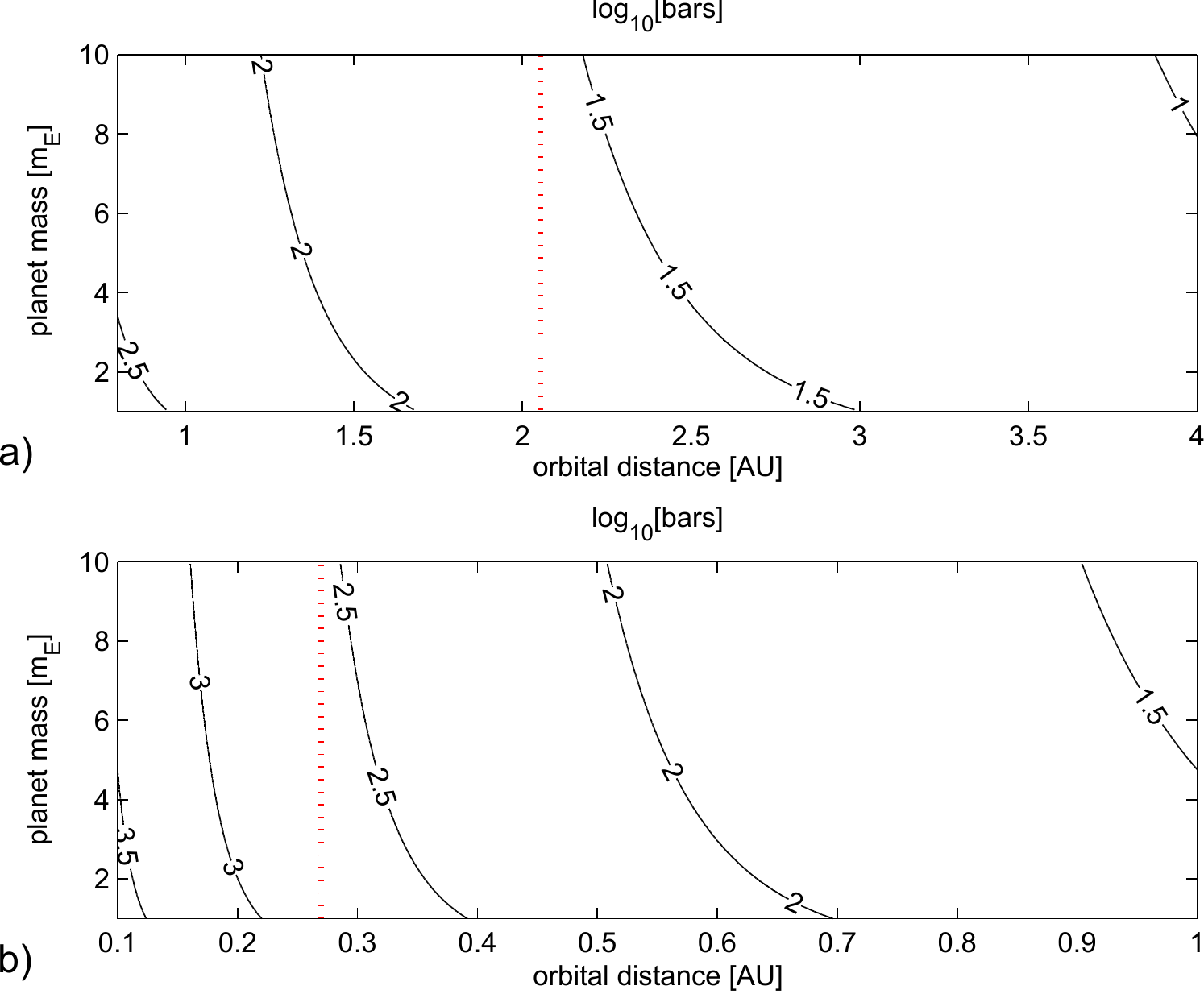}}
	\end{center}
	\caption{Total amount of hydrogen lost due to XUV energy-limited / Jeans thermal escape as a function of planet mass and orbital distance a) in a G-class system after 100 MYr and b) in an M-class system after 1 GYr. For reference, the dotted red line shows the critical H$_3^+$ cooling limit of \cite{Koskinen2007} for Jupiter-mass planets under saturated XUV conditions.}\label{totalescape}
\end{figure}

Figure \ref{totalescape} shows the total quantity of hydrogen that can be lost to space in time $t_{XUV}$ as a function of planet mass and orbital distance for G- and M-class stars. Due to the relatively small radii of rocky planets and high exospheric temperatures under elevated XUV conditions, energy-limited loss governed by (\ref{eq:elim}) occurs instead of Jeans escape almost everywhere. As can be seen, more than 100 bars of hydrogen may be lost in G-class systems for orbits closer than 1.5 AU, while for M-class systems, the 100-bar boundary is 0.5-0.7 AU. Accounting for the slight difference in luminosities used, these results are in approximate agreement with those of \cite{Pierrehumbert2011}, although they predict less escape than that shown in Figure \ref{totalescape} for more distant orbits. \cite{Pierrehumbert2011} used a slightly more complex exosphere model that predicted the exospheric temperature numerically, but still neglected all significant radiative cooling effects. Onset of Jeans escape at lower values of $d$ and $M$ due to lower exospheric temperatures in their model is the most likely cause of the difference in the results.

\begin{figure}[h]
	\begin{center}
		{\includegraphics[width=5.5in]{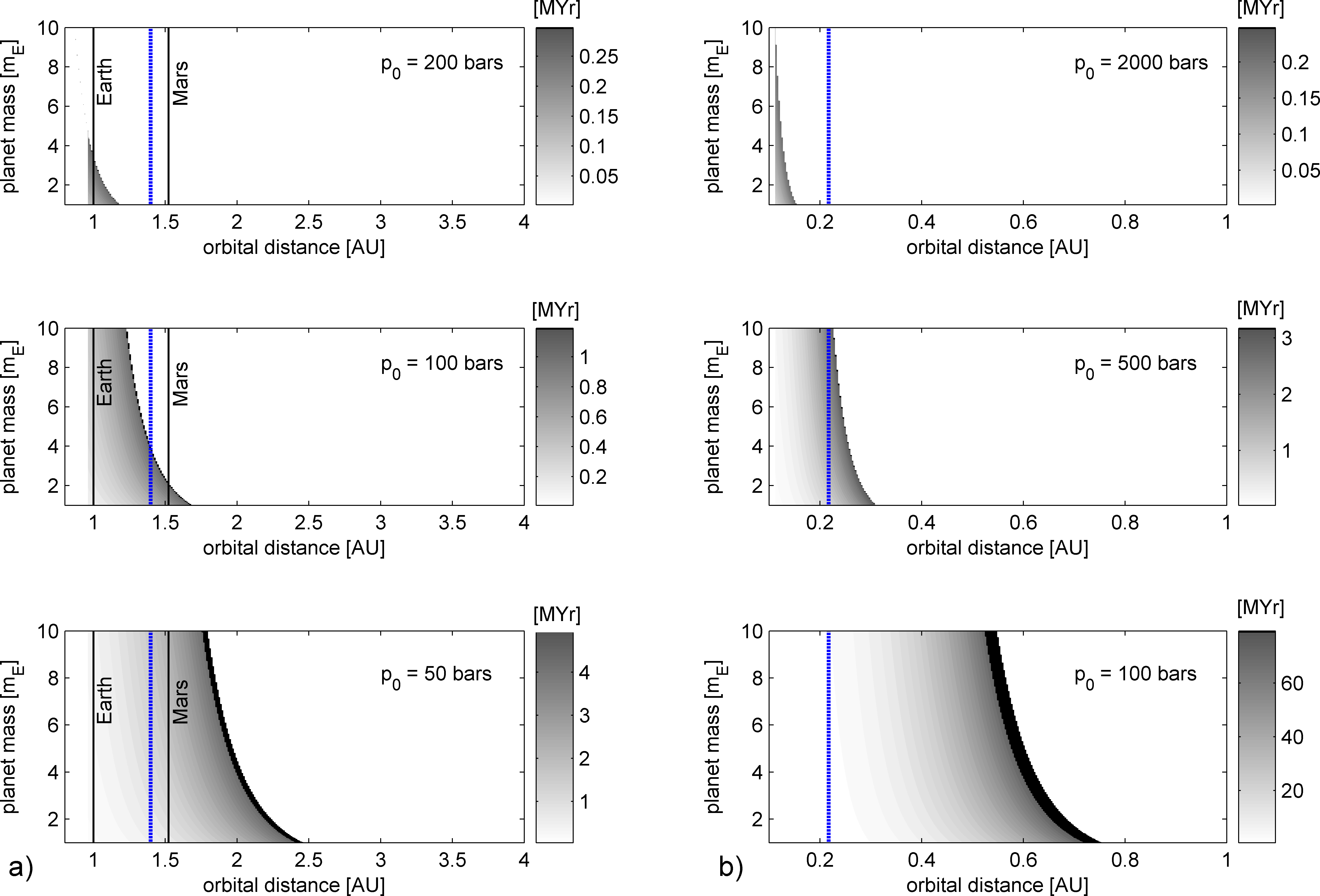}}
	\end{center}
	\caption{Duration of habitable conditions $\Delta t_{hab}$ as a function of planet mass and orbital distance for various starting atmospheric hydrogen inventories. a) and b) show results for G- and M-class systems, respectively. Results are not plotted for situations where insufficient hydrogen is removed to allow surface temperatures lower than $T_{max}=333$ K. The dashed blue line shows the outer edge of the CO$_2$-warming habitable zone for the given stellar luminosity, while in a) the thin solid lines show the orbits of Earth and Mars. The black region shows cases for which the atmosphere is eroded just the right amount to allow permanent surface temperatures in the 273 to 333 K range.}\label{timescales}
\end{figure}

Given our simplifying assumption of constant saturated XUV flux throughout the time period $t_{XUV}$, (\ref{eq:pressures}), (\ref{eq:elim}) and (\ref{eq:jeans}) can be combined to give an estimate of the duration of habitable conditions on the planet's surface
\begin{equation}\label{eq:durhab}
\Delta t_{hab} = [{p(T_{max})-p(T_{min})}]\slash{\phi}.
\end{equation}
Figure \ref{timescales} shows $\Delta t_{hab}$ in G- and M-class systems, assuming various initial H$_2$ inventories $p_0$. In situations where there is too little atmospheric loss to reach $p(T_{max})$ after $t_{XUV}$, the results have not been plotted. As can be seen, transient habitable conditions occur for many values of $d$, $M$ and $p_0$. The habitable duration $\Delta t_{hab}$ varies widely, from a few thousand years for close-in planets to values approaching $t_{XUV}$ for more distant ones. In some cases, the amount of atmospheric erosion is just right to leave the planet with surface temperatures in the habitable range permanently (black regions in Fig. \ref{timescales}). However, these cases represent only a small fraction of the total.

The starting H$_2$ pressure is a critical factor in the calculation. As Figure \ref{timescales} shows, for large values of $p_0$ only planets close to the star experience transient habitable periods. \cite{Pierrehumbert2011} argue based on the results of \cite{Rafikov2006} that initial H$_2$ pressures of 1-100 bar are possible, although it is likely that the exact amount in any situation will depend sensitively on the nature of accretion in the final stages of planet formation. As M-stars tend to have elevated XUV levels for longer periods than G-stars, planets around them can lose larger quantities of hydrogen, and  hence have a greater chance of experiencing transient hydrogen-induced habitable periods given starting H$_2$ pressures of greater than a few tens of bars.

The influence of exospheric radiative cooling on escape were neglected in this analysis for simplicity, but it could nonetheless be important in some cases. The effect of H$_3^+$ cooling can be crudely investigated by fixing the exospheric temperature at a constant value for XUV levels below the critical threshold $F_{XUV}^{crit}\sim 0.18$ W m$^{-2}$ given by \cite{Koskinen2007}. If we assume \mbox{$T_{exo}=3000$~K} for all planets outside a radius $d_{crit}=\sqrt{10^{-3.1}L \slash 4\pi F_{XUV}^{crit}}$, based on their Fig. 1, $\lambda \leq 2.8$ then requires $M \leq 1.16 M_{\oplus}$. For larger planetary masses, inefficient Jeans escape dominates and the amount of hydrogen removed from the atmosphere is significantly reduced. Unfortunately, \cite{Koskinen2007} examined only Jupiter-mass planets around Sun-like stars in their study, so we are unable to investigate the scaling of $T_{exo}$ with planetary mass and star type here. However, if H$_3^+$ cooling effects for terrestrial-mass planets are as strong as those for gas giants, transient habitable periods may be less common outside the critical radius than Fig. \ref{timescales} suggests.

\begin{figure}[h]
	\begin{center}
		{\includegraphics[width=4in]{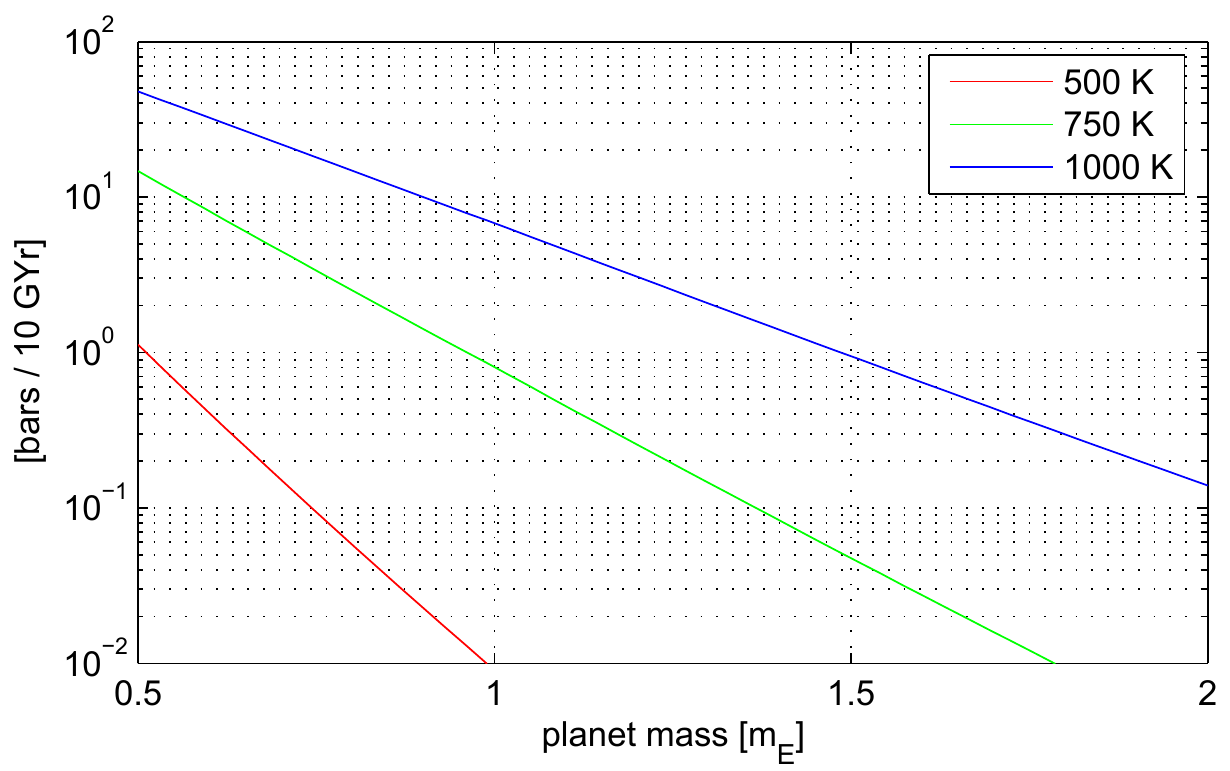}}
	\end{center}
	\caption{
Total amount of hydrogen lost due to atmospheric Jeans escape only after \mbox{10 GYr} as a function of planet mass, for planets with constant exospheric temperatures. For these relatively low values of $T_{exo}$, efficient hydrogen dissociation was no longer assumed, and the Jeans flux (\ref{eq:jeans}) was calculated using the collision cross-section for molecular hydrogen $\sigma_{H_2} = \pi (1.2\times10^{-10})^2$ m$^2$, to give a simple lower limit on the total escape.}\label{postXUVloss}
\end{figure}

For low-mass planets, thermal escape processes that do not depend on the stellar XUV flux may also contribute to H$_2$ loss and hence eventual transient habitability. In the Solar System, the variation in mean exospheric temperatures for the gas giants is surprisingly small, with typical values in the range $700-1000$ K despite the large differences in solar XUV insolation. Internal processes such as heating due to magnetic field ion acceleration and gravity wave breaking are believed to dominate the energy budget \citep{Lunine1993}. If similar mechanisms were present for rocky exoplanets with hydrogen envelopes, they would continue to lose gas to space even after the initial phase of saturated stellar XUV fluxes. Figure \ref{postXUVloss} shows the amount of hydrogen lost after 10 GYr due to atmospheric Jeans escape only, for planets of different masses and fixed $T_{exo}$ values. As can be seen, planets of Earth mass or lower can continue to lose significant amounts of hydrogen even far from their host star if their exospheric temperatures remain above $\sim$750 K. Such bodies will of course be difficult to observe, but the example serves to illustrate the variety of situations that may lead to transient habitable scenarios.

\section{Discussion}
In summary, if a planet with equilibrium temperature lower than that of Earth has multiple bars of hydrogen in its atmosphere early in its evolution (either due to direct capture from the nebula or outgassing after accretion) but subsequently loses it due to escape, it will almost certainly pass through a transient state where liquid water can form on its surface. It has been shown that the duration of habitable conditions can range from thousands to hundreds of millions of years, although planets left with exactly the right amount of hydrogen for permanent habitable conditions only occur in rare cases.

Several effects that we have not considered could alter the simple analysis presented here. For one, we have assumed a short, discrete phase of planetary accretion, but meteorite impacts can continue long after the main stages of formation are over. Events such as the Late Heavy Bombardment in the Solar System could erode primordial hydrogen atmospheres and evaporate oceans, potentially destroying any emerging life. However, as noted by \cite{Abramov2009}, even on an Earth-mass planet relatively large impactors ($r\geq50$ km) are incapable of sterilizing the entire surface.

Perhaps more restrictive is the fact that we have only described the evolution of pure H$_2$ atmospheres. For more realistic hydrogen-dominated atmospheres containing a mixture of gases, other sources of opacity will be present that could increase greenhouse warming. In addition, photochemistry in the planet's upper atmosphere can lead to the formation of aerosols and organic hazes, which can increase the planetary albedo or cause `anti-greenhouse' effects \citep[e.g., ][]{McKay1991}. These processes could quantitatively alter the results described here, and should be investigated in future using more detailed models. Nonetheless, it is difficult to see how they could invalidate the basic mechanism entirely. The effectiveness of hydrogen as a greenhouse gas means that planets with primordial H$_2$ envelopes should have high initial surface temperatures even if they form far from the host star. In cases where the hydrogen envelope is later removed and the planets receive significantly less starlight than Earth, transient periods of habitable surface temperatures of some duration should therefore always result.

The reducing chemistry of hydrogen-rich atmospheres provides fertile conditions for the formation of some of the basic building blocks of life, such as amino acids and ribonucleotides \citep{Miller1953,Ferris1978,Powner2009}. Provided C, N and O-bearing species are also present, an exoplanet with a slowly evaporating hydrogen atmosphere would likely therefore experience a steady rain of organic compounds towards the surface due to photolysis in the upper atmosphere. If the atmosphere were dense enough to cause high temperatures at the surface, these compounds would be destroyed by thermolysis relatively rapidly. However, as the greenhouse effect of hydrogen diminished, temperatures would decrease, allowing them to begin to accumulating on the surface. Moderately high temperatures (60 - 100 $^\circ$C) dramatically accelerate key reactions in primordial biochemistry \citep{Stockbridge2010}, so the initial stages of a planet's transiently habitable period might provide particularly favourable conditions for the development of complex reaction pathways on the surface.

In an environment dominated by hydrogen-bearing species, the efficiency of redox reactions available for metabolism would clearly be very different from that of either present-day Earth or the weakly reducing N$_2$-CO$_2$-H$_2$O atmospheres generally believed to be representative of the Archean \citep[e.g. ][]{Kasting1993b}. Nonetheless, emergent organisms could potentially extract energy from a variety of mechanisms. Volcanic emission of gases such as CO$_2$ or sulphate aerosols from an oxidising mantle \citep{ElkinsTanton2008} would allow for some of the redox reactions already known to be used by methanogens and sulphur-reducing bacteria on Earth. Photochemistry itself provides further possibilities: by definition it is driven by the free energy from photons and as such results in compounds that are out of equilibrium with their environment \citep{Pascal2011}.  To give one example, dissociation of water vapour and the subsequent loss of hydrogen to space results in highly reactive radicals such as OH*, which can cause the formation of oxidized trace species (e.g. H$_2$O$_2$) that are transported to the surface via rainout.

After the loss of the hydrogen envelope in these scenarios, heavier gases such as methane or carbon dioxide could remain in the atmosphere, depending on their volatility and escape rates. If these gases provided insufficient greenhouse warming to keep temperatures above 0 $^\circ$C, surface liquid water would no longer be possible globally, although processes such as vulcanism and geothermal heating would still allow locally habitable regions and hence some further opportunity for life to develop. Investigating the end state of atmospheric evolution self-consistently in these cases, while challenging, would be an interesting subject for future research.

If life readily emerges under these conditions but cannot adapt to the changing climate by altering either itself or its environment, transient hydrogen greenhouse warming may be of little relevance to the search for life around other stars. In this case, constraints based on abiotic climate stabilization mechanisms such as the carbonate-silicate cycle \citep{Walker1981} will probably determine the frequency of life-supporting planets. The carbonate-silicate cycle requires plate tectonics for CO$_2$ recycling, and the extent to which this phenomenon depends on a planet's mass, internal heat sources and mantle composition is still under investigation \citep{Valencia2007,Korenaga2010}. If only a sub-set of rocky planets in the CO$_2$ habitable zone have active carbonate-silicate cycles, the prevalence of life-supporting planets in the stellar neighborhood could therefore be small.

If, however, the adaptability of life to diverse habitats through evolution and environmental modification is high, the frequency of conditions allowing its formation may be the critical factor. Life on Earth is adapted to a wide range of physical and chemical environments, and has colonized all regions of the planet where even minimal liquid water is present, including the Antarctic subsurface \citep{Cary2010}. It has been shown elsewhere \citep{Laughlin2000,Abbot2011} that even planets receiving essentially no stellar flux are able to maintain ice-covered liquid oceans through geothermal heating alone. Hence even in the absence of other warming processes, localized habitable environments would remain in the time after transient hydrogen warming had ceased.

Since the Archean era, life has also extensively modified the composition of the atmosphere. The extent to which this modification has regulated surface temperatures to optimize planetary habitability remains a subject of intense debate \cite[e.g., ][]{Lovelock1974,Kirchner2003}, and resolving the argument has proved difficult, in part due to our inability to study the Earth's atmosphere in different epochs directly. However, it has been shown here that appropriate conditions for the origin of life may occur in a fairly wide range of cases. Hence if living ecosystems are capable of regulating global climate, even planets outside the standard CO$_2$-warming habitable zone might retain biospheres for longer periods. The boundaries of the CO$_2$ habitable zone are fairly well-defined as a function of star type, and global biospheres may be detectable from e.g., the presence of non-equilibrium gases in a planet's atmosphere. Given the planned future searches for nearby planets with biosignatures, therefore, in theory it could eventually be possible to assess the relative importance of abiotic vs. biotic climate stabilization processes directly.

\ack This article has benefited from fruitful discussions with Franck Selsis, Fran\c cois Forget, Robert Pascal and Lee Grenfell.


\bibliographystyle{plainnat}

\label{lastpage}

\end{document}